\def\targetformat{arXiv}		

\ifx\targetformat\undefined	
\def\clsstyle{prl} 

\newcommand{\appref}[1]{the Supplemental Materials}
\newcommand{\arxivtext}[1]{}
\newcommand{\prltext}[1]{#1}
\else		
\def\clsstyle{pra} 

\newcommand{\appref}[1]{App.~\ref{#1}}
\newcommand{\arxivtext}[1]{#1}
\newcommand{\prltext}[1]{}

\fi

\documentclass[aps,\clsstyle,reprint,twocolumn,showpacs,superscriptaddress,footnoteinbib]{revtex4-1}

\usepackage{amsmath, amssymb, bm, braket, dsfont, times, amsthm, graphicx, xfrac}
\usepackage[breaklinks,colorlinks,allcolors=blue]{hyperref}

\begin{document}
\title{Stable Luttinger liquids and emergent $U(1)$ symmetry in constrained quantum chains}

\author{Ruben Verresen}
\affiliation{Max-Planck-Institute for the Physics of Complex Systems, 01187 Dresden, Germany}
\affiliation{Department of Physics, T42, Technische Universit\"at M\"unchen, 85748 Garching, Germany}
\author{Ashvin Vishwanath}
\affiliation{Department of Physics, Harvard University, Cambridge MA 02138, USA}
\author{Frank Pollmann}
\affiliation{Department of Physics, T42, Technische Universit\"at M\"unchen, 85748 Garching, Germany}
\affiliation{Munich Center for Quantum Science and Technology (MCQST), 80799 Munich, Germany}

\date{\today}                                       

\begin{abstract}
	
We explore the effect of local constraints on one-dimensional bosonic and fermionic ground state phases. Motivated by recent experiments on Rydberg chains, we constrain the occupation of neighboring sites in known phases of matter. Starting from Kitaev's topological superconductor wire, we find that a soft constraint induces a stable gapless Luttinger liquid phase. While Luttinger and Fermi liquids are usually unstable to superconducting proximity effects, the constraint suppresses pair creations, allowing for an emergent $U(1)$ symmetry and gaplessness. We substantiate this intuitive picture using field theoretical and Bethe ansatz methods. In particular, in the hard constraint limit, the model is explicitly $U(1)$-symmetric and integrable. For the corresponding spin-$1/2$ chains related by a Jordan-Wigner transformation, the Luttinger liquid is stabilized by the $\mathbb Z_2$ spin flip symmetry. Longer-range constraints stabilize gapless phases even without $\mathbb Z_2$ symmetry, connecting to the seminal work by Fendley, Sengupta and Sachdev [Physical Review B 69, 075106 (2004)], clarifying how the gapless floating phase observed therein can be vastly extended.
\end{abstract}
\maketitle

\textbf{\emph{Introduction.}}---Locality is a basic tenet of many-body quantum theory, often accompanied by the Hilbert space being a tensor product of smaller spaces (e.g., qubits). Imposing local constraints gives rise to low-energy \emph{effective} Hilbert spaces which need \emph{not} have a tensor product structure, and such constrained models have a rich history. The archetypal example is water ice \cite{Bernal33,Pauling35,Lieb67}, with generalizations to ice-type/vertex models \cite{Baxter82} and dimer models \cite{Anderson87,Rokhsar88,Moessner08}---leading to, e.g., emergent magnetic monopoles \cite{Castelnovo08}. This is a broad field including string-nets \cite{Freedman04,Levin05} and lattice gauge theories \cite{Kogut79}.

In recent years, there has been a surge of interest in the one-dimensional setting with quantum simulators made of cold Rydberg atoms \cite{Jaksch00,Weimer10,Viteau11,Bernien17}. These chains exhibit the Rydberg blockade \cite{Lukin01,Urban09,Gaetan09}: two neighboring sites cannot both be excited. Quantum scars are an interesting dynamical phenomenon to have emerged from this \cite{Turner17,Turner18,Choi18}. Previous studies of ground state phase diagrams focused on the interplay between a field and the constraining terms \cite{Fendley04,Lesanovsky11,Lesanovsky12,Lesanovsky12b,Chepiga19,Giudici19}.

Our motivation is to study the possible structure of general ground state phase diagrams. In particular, what phases can be realized given certain symmetries? How to characterize them? Is the classification in terms of symmetry-breaking and symmetry-protected topological (SPT) phases? Such questions have only recently been answered in the unconstrained case \cite{Schnyder08,Kitaev09,Ryu10,Fidkowski11class,Turner11class,Chen10,Schuch11}. This is a wide and ambitious program. In this work, we focus on a curious aspect of such phase diagrams, namely gapless phases which are unusually stable.

To appreciate the precarious nature of gaplessness, consider $H = - \sum_i \left( c_i^\dagger c_{i+1} + \textrm{h.c.} \right)$, the simplest Luttinger liquid (i.e.~a one-dimensional Fermi liquid). Its dispersion is $\varepsilon_k = -2\cos k$, hence its Fermi surface/points are at $k_F = \pm \pi/2$. This model is susceptible to so-called charge-density-wave (CDW) and superconducting (SC) instabilities. These involve the repulsion between two bands, opening up a gap at the Fermi points. The CDW instability entails breaking translation symmetry, allowing $c_k$ to couple to $c_{k+\pi}$. The SC instability entails breaking $U(1)$ symmetry, allowing particles ($c^\dagger_k$) to couple to holes ($c_{-k}$). We hence need translation and $U(1)$ symmetries to stabilize the gapless phase. Note that the SC instability leads to the well-known Kitaev chain \cite{Kitaev01}:
\begin{equation}
H_K = - \sum_i \left( c_i^\dagger c^{\vphantom \dagger}_{i+1} + c_i^\dagger c_{i+1}^\dagger + \textrm{h.c.} \right).
\end{equation}
This Hamiltonian forms a gapped topological phase protected by the $\mathbb Z_2$ fermionic parity symmetry and consequently exhibits Majorana edge modes \footnote{Note that $\gamma_L \propto c_1 + c_1^\dagger$ commutes with $H_K$ for open boundaries.}.

In our work, we add the soft (`Rydberg') constraint $U \sum_i n_i n_{i+1}$ to the Kitaev chain. For large $U$, a stable Luttinger liquid (LL) emerges despite the Hamiltonian not being $U(1)$-symmetric. This can be understood intuitively: if the density of particles is not too dilute, pair fluctuations come with a penalty cost $\sim U$; their suppression leads to an emergent $U(1)$ symmetry at low energies. More quantitatively, in the strong-coupling limit $U \to \infty$, the effective Hamiltonian in the constrained Hilbert space coincides with a known $U(1)$-symmetric, integrable model \cite{Alcaraz99}. Its field theory is a LL with parameter $K < 1/2$, implying stability against superconducting perturbations. In addition, an incommensurate Fermi momentum avoids the CDW instability. At finite $U$, the non-integrable model can be seen as a perturbation of a nearby $U(1)$-symmetric, integrable model, giving semi-quantitative insight into the stability of the emergent gapless phase.

While we focus on the fermionic context, particle statistics are not a key player in one dimension: the Jordan-Wigner transformation allows to equivalently discuss spin-$1/2$ models (listed in Appendix~\ref{app:spin}). One key difference in the spin chain language is that the $\mathbb Z_2$ spin flip symmetry---which corresponds to the fermionic parity symmetry stabilizing the LL---can be broken (either spontaneously or explicitly). However, we show that longer-range constraints give rise to regions where the LL parameter $K<1/8$, implying stability \emph{without} $\mathbb Z_2$ symmetry. For constraints on nearest and next-to-nearest neighbors, the phase is very narrow in the phase diagram, consistent with observations by Fendley, Sengupta and Sachdev \cite{Fendley04}. Longer-range constraints vastly stabilize the LL.

In addition to showing the ubiquity of stable Luttinger liquids in constrained systems, we touch upon the issue of what other phases of matter can be realized in constrained Hilbert spaces. In particular, we numerically study the effect of terms that break $U(1)$ and translation symmetry. For weak perturbations, we confirm the stability of the LL. For strong perturbations, various distinct topological and symmetry-breaking phases emerge.

\begin{figure}
\includegraphics[scale=0.98,trim=0.1cm 0.1cm 0.15cm 0.1cm,clip]{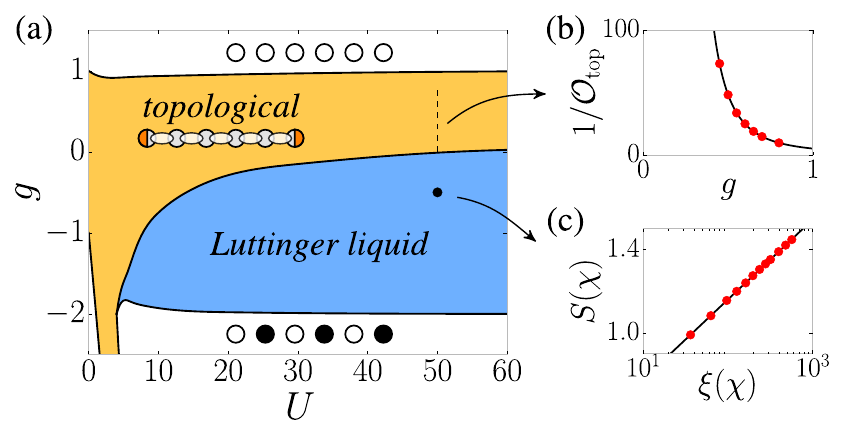}
\caption{(a) Phase diagram for the Kitaev chain with a soft constraint, Eq.~\eqref{eq:Ham1}, exhibiting a stable Luttinger liquid (LL) phase. Between the topological and LL phase, there is a BKT transition. (b) This is confirmed along the dashed line ($U=50$): the topological order parameter $\mathcal O_\textrm{top} \equiv \lim_{|i-j|\to\infty} \big| \langle (c_i-c_i^\dagger) \exp\big( i \pi \sum_{i<k<j} n_k \big) (c_j + c_j^\dagger) \rangle \big|$ goes to zero as $\exp\left(- a/\sqrt{g-g_0} \right)$; fitting locates the transition at $g_0 \approx 0$.
(c) In the LL phase, we can perform entanglement scaling $S(\chi) \sim \frac{c}{6} \ln \xi (\chi)$ to determine the central charge \cite{Calabrese04,Pollmann09}; e.g. at $g=-\sfrac{1}{2}$ and $U=50$, fitting gives $c = 1.000 \pm 0.002$. \label{fig:phasediagramIsing}}
\end{figure}

\textbf{\emph{Constraining the Kitaev and Ising chain.}}---Starting from the Kitaev chain with a chemical potential, we energetically introduce a soft constraint:
\begin{equation}
H_\textrm{soft} = H_K + 2g \sum_i n_i + U \sum_i n_i n_{i+1}. \label{eq:Ham1}
\end{equation}

Figure \ref{fig:phasediagramIsing} shows the phase diagram obtained with the infinite density matrix renormalization group (iDMRG) method \cite{White92,Kjaell13}. For $U=0$, we have Ising critical points at $g=\pm 1$, beyond which there are trivial gapped phases.

For finite $U$, we see a large gapless phase which is entered by a Berezinski{\v i}-Kosterlitz-Thouless (BKT) transition \cite{Berezinskii72,Kosterlitz73}. The central charge, $c=1$, is consistent with this being a Luttinger liquid (LL). This is surprising since LLs usally appear in the presence of stabilizing symmetries, such as $U(1)$, or at fine-tuned critical points. To better understand this phase, we study the $U \to \infty$ limit in the next section.

We mention in passing that the spin-$\sfrac{1}{2}$ chain corresponding to $H_\textrm{soft}$ is the transverse-field Ising chain with a soft constraint. Interestingly, using a nonlocal (Kramers-Wannier) transformation, this spin chain can be mapped onto the axial/anisotropic next-nearest neighbor Ising (ANNNI) chain. This mapping is also nonlocal in parameter space, e.g. mapping the $U \to \infty$ limit onto a single (`multiphase') point. Hence, $H_\textrm{soft}$ puts the ANNNI model in a new light.
For example, $H_\textrm{soft}$ having a gapless phase for $g=0$ (numerically observed in Fig.~\ref{fig:phasediagramIsing} and analytically argued in the following section) is equivalent to the gapless phase of the ANNNI mode having finite extent above the multiphase point, which had been debated in the literature; details are discussed in Appendix~\ref{app:ANNNI}.

\begin{figure}
\includegraphics[scale=0.98,trim=0.1cm 0.1cm 0.2cm 0.1cm,clip]{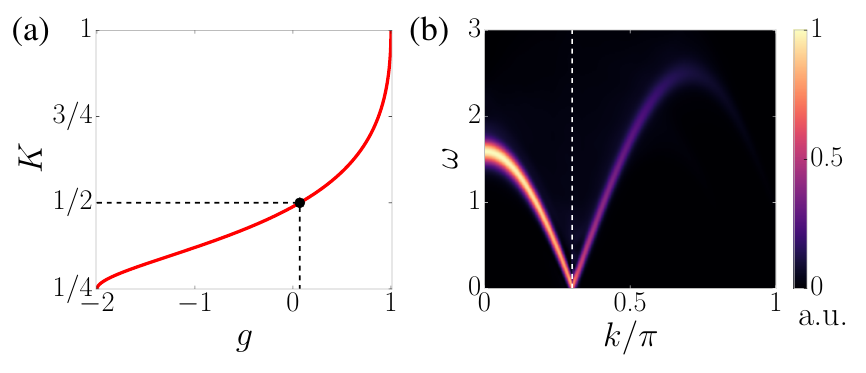}
\caption{Absence of SC and CDW instabilities for the Kitaev chain with a hard constraint, Eq.~\eqref{eq:Hamconstrained} (i.e. the $U\to \infty$ limit of Fig.~\ref{fig:phasediagramIsing}). (a) The Luttinger liquid parameter $K$ from the Bethe ansatz solution. For $-2 \leq g \lessapprox 0.0697$, the model is critical with $K < \sfrac{1}{2}$, implying stability against explicitly breaking $U(1)$ symmetry. (b) The (fermionic) spectral function for $g=0$. The gapless quasiparticles are at the \emph{incommensurate} wave vector $k_F/\pi = 1-\sqrt{K} \approx 0.301 $ (white dashed line), guaranteeing stability against dimerization. \label{fig:PXYP}}
\end{figure}

\textbf{\emph{Stability and $\bm{U(1)}$ symmetry in the constrained limit.}}---In the strong-coupling limit $U\to \infty$, the effective Hilbert space consists of all states where neighboring sites cannot both be occupied. This is an example of a Hilbert space without a tensor product structure, e.g. evidenced by its dimension asymptotically scaling as $\sim \varphi^N$ (with $\varphi$ being the golden ratio) \cite{Moessner01}. At leading order, the effective Hamiltonian is given by Eq.~\eqref{eq:Ham1} projected into this subspace, i.e. $H_\textrm{hard} = \mathcal P H_\textrm{soft} \mathcal P$:
\begin{align}
H_\textrm{hard} &= - \sum_i P_{i-1} \left( c_i^\dagger c^{\vphantom \dagger}_{i+1} + \textrm{h.c.} \right) P_{i+2} + 2 g \sum_i n_i, \label{eq:Hamconstrained}
\end{align}
where $P_i = 1-n_i$ projects onto an empty site. (This may be dubbed the PXXP model, akin to the PXP model \cite{Turner17,Turner18,Choi18}.) The superconducting terms have disappeared! This is because they would violate the constraint. Remarkably, this model is known to be integrable and a LL for $-2 \leq g \leq 1$ \cite{Alcaraz99}.

At next order in perturbation theory, the $U(1)$ symmetry of the Hamiltonian is broken by second-nearest neighbor couplings: $- \frac{2}{U} \sum_i  \left( P_{i-2} c^\dagger_{i-1}  P_i  c^\dagger_{i+1} P_{i+2} + \textrm{h.c.} \right)$ (see Appendix~\ref{app:pert} for the complete effective Hamiltonian at this order). However, we now show that these terms do \emph{not} destroy the $U(1)$ symmetry at low energies: the gapless phase is robust for $-2 \leq g \lessapprox 0.0697$.

This stability can be understood from the Luttinger liquid parameter $K$. A defining characteristic of $K$ is that it is the inverse of the scaling dimension of pair creation. Hence, if $K<1/2$, superconducting fluctuations have a scaling dimension larger than two and are thus irrelevant in the renormalization group flow. From the Bethe ansatz solution of Eq.~\eqref{eq:Hamconstrained}, $K$ and $g$ are related as follows \cite{Alcaraz99} (see Appendix~\ref{app:XXZ2} for a derivation):
\begin{equation}
\pi g = \big( \sin(x) - x \cos(x) \big) \big|_{x = \pi / \sqrt{K}}. \label{eq:Bethe}
\end{equation}
This is plotted in Fig.~\ref{fig:PXYP}(a), with $K$ evolving from $1/4$ to $1$. For $-2 \leq g < \frac{1}{\pi}\sin(\sqrt{2}\pi) - \sqrt{2}\cos(\sqrt{2}\pi) \approx 0.0697$, we have $1/4 \leq K < 1/2$, avoiding the SC instability.

The model also avoids the CDW instability. It can be shown that $K = (1-\rho_0)^2$, where $\rho_0$ is the ground state filling \cite{Alcaraz99} (see also Appendix~\ref{app:XXZ2}). Hence, Eq.~\eqref{eq:Bethe} tells us how $g$ determines the filling. In the limiting cases $g=-2$ and $g=1$, the filling is $\rho_0 = 1/2$ and $\rho_0 = 0$, respectively. In between, it takes all intermediate values. This implies that the filling and the Fermi point $k_F/\pi = \rho_0$ are generically incommensurate. Hence, no matter how many times we fold the BZ, the gapless mode cannot gap out; all we need is translation symmetry with respect to \emph{some} unit cell. We confirm the incommensurability of $k_F$ in Fig.~\ref{fig:PXYP}(b), showing the fermionic spectral function $\mathcal A(k,\omega) = - \frac{1}{\pi} \textrm{Im} G^R(k,\omega+0^+)$ for $g=0$.
This is obtained by calculating the dynamical correlation function using DMRG \cite{White08,Zaletel15}.

In summary, the effective model \eqref{eq:Hamconstrained} is protected against the SC instability due to $K < \frac{1}{2}$ and against the CDW instability due to incommensurability. Both properties are conceptually related to the constraint: the former due to the suppression of pair fluctuations, the latter due to the non-trivial optimization of ground state filling.

\begin{figure}
\includegraphics[scale=1,trim=0.25cm 0.25cm 0.15cm 0.2cm,clip]{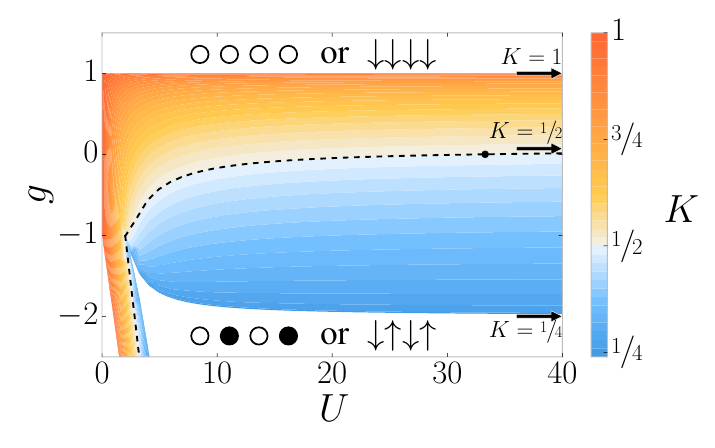}
\caption{Phase diagram and Luttinger liquid parameter $K$ for the integrable model $H_\textrm{soft}^{U(1)}$. The dashed line separates the region where $K\leq \sfrac{1}{2}$. The black dot denotes where this line crosses $g=0$. The three arrows indicate values of $K$ in the constrained limit $U \to \infty$ (for $g= -2$, $0.0697$ and $1$; see Fig.~\ref{fig:PXYP}). The region with $K\leq \sfrac{1}{2}$ is stable against explicitly breaking $U(1)$, which qualitatively explains the Luttinger liquid found in the non-integrable model (see Fig.~\ref{fig:phasediagramIsing}). \label{fig:phasediagramXY}}
\end{figure}

\textbf{\emph{A nearby integrable model with soft constraint.}}---In the previous section, we saw that in the $U\to \infty$ limit of Eq.~\eqref{eq:Ham1}, the strongly-interacting Kitaev chain becomes integrable and explicitly $U(1)$-symmetric. This gave us insight into the stability of the gapless phase. We now point out that although the Kitaev chain with a \emph{soft} constraint is non-integrable, it can be seen as a perturbation of an integrable, $U(1)$-symmetric model. We use this to qualitatively explain the phase diagram obtained in Fig.~\ref{fig:phasediagramIsing}.
Note that if we remove the SC terms of Eq.~\eqref{eq:Ham1}, we obtain
\begin{equation}
H_\textrm{soft}^{U(1)} = \sum_i \left( - c^\dagger_i c^{\vphantom \dagger}_{i+1} -  c_{i+1}^\dagger c^{\vphantom \dagger}_i + 2g n_i + U n_i n_{i+1} \right). \label{eq:U1soft}
\end{equation}
In the constrained limit $U \to \infty$, this leads to the \emph{same} effective Hamiltonian as before! In other words, $H_\textrm{soft} \approx H_\textrm{soft}^{U(1)}$ for large $U$. In particular, whenever $H_\textrm{soft}^{U(1)}$ is a Luttinger liquid phase with $K<\sfrac{1}{2}$ (with appreciable values of $U$), we can expect $H_\textrm{soft}$ to be critical.

This relationship is useful since $H_\textrm{soft}^{U(1)}$ is integrable.
In particular, the Jordan-Wigner transformation maps it to the paradigmatic XXZ chain in a field. Ref.~\cite{Bogoliubov86} presents the integral equations for the Luttinger liquid parameter of the XXZ chain. Numerically solving these equations, we arrive at the phase diagram in Fig.~\ref{fig:phasediagramXY} (see Appendix~\ref{app:XXZ} for details). If we perturb $H^{U(1)}_\textrm{soft}$ with $U(1)$-breaking terms, we expect the region with $K<\sfrac{1}{2}$ to remain gapless. This indeed qualitatively matches the gapless phase of $H_\textrm{soft}$ in Fig.~\ref{fig:phasediagramIsing}, whereas the $K>\sfrac{1}{2}$ region has flown to the gapped topological Kitaev phase. 

\begin{figure}
\includegraphics[scale=1,trim=0.15cm 0.15cm 0.1cm 0.2cm,clip]{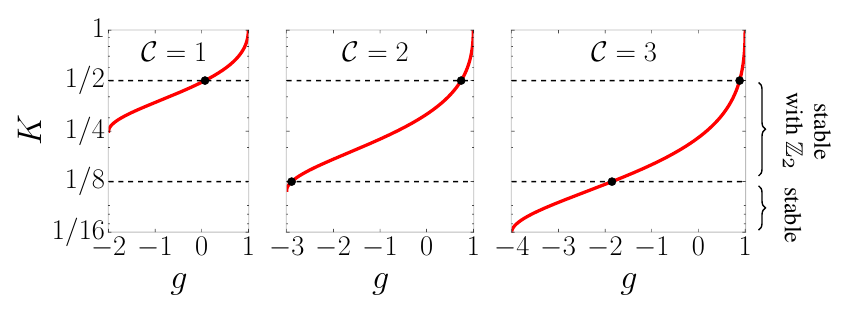}
\caption{Stronger constraints imply stronger stability. Shown is the Luttinger liquid parameter for the constrained model in Eq.~\eqref{eq:stronger} where sites within a distance $\mathcal{C}$ cannot both be occupied. If $K<1/2$, the LL is stable as long as we preserve parity symmetry. For $K<1/8$ (occurring for $\mathcal C\geq 2$) it is stable even upon explicitly breaking $\mathbb Z_2$ parity symmetry in hardcore boson/spin chains. \label{fig:C}}
\end{figure}

\textbf{\emph{Stability in the spin chain language through longer-range constraints.}}---The above demonstrated the stability of LLs in fermionic chains. However, in the hardcore boson/spin chains obtained by a Jordan-Wigner transformation, \emph{single}-particle fluctuations can occur. These are considerably more effective at opening gaps: their scaling dimension is $\sfrac{1}{4K}$, which is relevant whenever $K>\sfrac{1}{8}$. There are two ways out of this instability: either we enforce the $\mathbb Z_2$ parity symmetry by hand, or we introduce stronger constraints; we explore the latter.

A natural way of suppressing single-particle fluctuations is by penalizing two particles within a distance $\mathcal C$ (until now $\mathcal C = 1$). One can then think of each particle as effectively extending over $\mathcal C+1$ sites. If the density is not too dilute, there will be no more room for single-particle fluctuations. Such generalized constraints are softly implemented by the term $U \sum_i \sum_{r=1}^{\mathcal C} n_i n_{i+r}$. Adding this to the Kitaev or Ising chain and taking $U\to \infty$ gives us a generalization of Eq.~\eqref{eq:Hamconstrained}. In the spin chain language, we obtain
\begin{equation}
H_\textrm{hard}^{(\mathcal C)} = \frac{1}{2} \sum_i P_{i-\mathcal C}^{(\mathcal C)} (X_i X_{i+1} + Y_i Y_{i+1}) P_{i+2}^{(\mathcal C)} + g \sum_i Z_i, \label{eq:stronger} 
\end{equation}
where the projector $P_i^{(\mathcal C)} = P_i \cdots P_{i+\mathcal C-1}$ ensures that we remain in the constrained Hilbert space.

Remarkably, this model is integrable for any value of $\mathcal C$ \cite{Alcaraz99}. It is a LL for $-(1 +\mathcal C) \leq g \leq 1$, attaining its maximal filling $\rho_0^\textrm{max} = \frac{1}{1+\mathcal C}$ at $g=-(1+\mathcal C)$. This signals a Pokrovsky-Talapov (PT) transition into a commensurate gapped phase \cite{Pokrovsky80}, and as predicted by PT universality, the Luttinger liquid parameter is $K_\textrm{PT} = \left(\rho_0^\textrm{max}\right)^2$ in this limiting case \footnote{More generally, one can prove that $K= (1- \mathcal C \rho_0)^2$ in this model. Note that $1- \mathcal C \rho_0^\textrm{max} = \rho_0^\textrm{max}$.}. Note that if $\mathcal C \geq 2$, then $K_\textrm{PT} < \sfrac{1}{8}$. More generally, one can derive the relationship between $K$ and $g$ \cite{Alcaraz99}:
\begin{equation}
\pi g = \mathcal C \left( x \cos\left( \pi( x - 1) / \mathcal C \right) -  \sin\left(\pi( x - 1) / \mathcal C  \right) \right) \big|_{x = 1/ \sqrt{K}}.
\end{equation}
(For $\mathcal C=1$, this reduces to Eq.~\eqref{eq:Bethe}.) This relationship is plotted in Fig.~\ref{fig:C}. We observe that for $\mathcal C \geq 2$, there are indeed gapless regions where $K < \sfrac{1}{8}$, ensuring stability against $\mathbb Z_2$-breaking terms.

For $\mathcal C= 2$, we can make a conceptual connection with the work of Fendley, Sengupta and Sachdev \cite{Fendley04}. They considered a model without on-site $\mathbb Z_2$ symmetry and found that when occupation was suppressed for both nearest and next-to-nearest neighbors, then there was a gapless phase of small but finite extent. This is qualitatively similar to Fig.~\ref{fig:C}, where there is a small region $-3 \leq g \lessapprox -2.90$ for which $K<1/8$ when $\mathcal C=2$. For $\mathcal C=3$, we see that this region has vastly increased to $-4 \leq g \lessapprox -1.85$ (moreover, $K < 1/2$ for $-4 \leq g \lessapprox 0.88$). Longer-range constraints clearly stabilize the Luttinger liquid against particle and pair fluctuations.

\begin{figure}
\includegraphics[scale=.975,trim=0.13cm 0.2cm 0.25cm 0.1cm,clip]{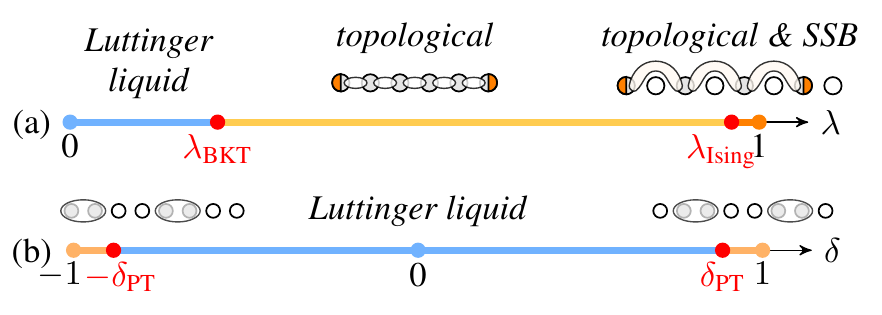}
\caption{Explicitly breaking $U(1)$ and translation symmetry in the constrained limit. (a) Phase diagram for $H_\lambda$ in Eq.~\eqref{eq:U1break} with $g=-1$. There is a BKT transition into a topological phase at $\lambda_\textrm{BKT} \approx 0.21$ and an Ising transition at $\lambda_\textrm{Ising} \approx 0.97$, beyond which translation symmetry is spontaneously broken. For $\lambda = 1$, we have the longer-range Kitaev chain in the constrained Hilbert space. (b) Phase diagram for $H_\delta$ in Eq.~\eqref{eq:dimer}. There is a Pokrovsky-Talapov transition at $\delta_\textrm{PT} \approx 0.884$ into a phase that spontaneously breaks the two-site translation symmetry. In addition to the sketched ground states, we also have the ones shifted by two lattice sites. For positive $\delta$, the ground states are SPTs with respect to the remaining inversion symmetry, but there are no physical edge modes. \label{fig:phasediagramconstrained}}
\end{figure}

\textbf{\emph{$\bm{U(1)}$-breaking and dimerization in the nearest-neighbor-constrained Hilbert space.}}---Above we have shown how constraints can stabilize a LL against perturbations that break $U(1)$, $\mathbb Z_2$ and translation symmetry. In this last part, we explore what phases are realized when making such perturbations strong. This touches upon the question of what phases of matter can be stabilized in constrained Hilbert spaces. As our starting point, we use $H_\textrm{hard}$ in Eq.~\eqref{eq:Hamconstrained} (i.e. $\mathcal C =1$).

In the unconstrained case, breaking the $U(1)$ symmetry of the fermionic chain yields a topological phase (a symmetry-breaking phase in the spin chain). Alternatively, introducing bond-alternation forms a symmetry-protected topological phase: the so-called Su-Schrieffer-Heeger chain in the fermionic case \cite{Su79} (the Haldane phase in the spin language with an appropriate choice of unit cell \cite{Haldane83,Hida92}). It is natural to wonder whether these phases are also realized in the constrained Hilbert space.

First, we perturb $H_\textrm{hard}$ with $U(1)$-breaking terms:
\begin{equation}
H_\lambda =  (1-\lambda) \; H_\textrm{hard}(g) + \lambda \; \mathcal P H'_K \mathcal P \label{eq:U1break}
\end{equation}
where $H'_K$ is the \emph{second}-nearest-neighbor Kitaev chain and $\mathcal P$ is the projector onto the constrained Hilbert space; such a term is perturbatively generated in $1/U$. 
The phase diagram obtained with iDMRG is shown in Fig.~\ref{fig:phasediagramconstrained}(a), where we chose $g=-1$. We indeed find that for large enough $\lambda$, the topological Kitaev chain phase is realized in the constrained Hilbert space. Interestingly, for even larger values of $\lambda$, the $\mathbb Z_2$ translation symmetry is spontaneously broken. (Numerical details for the critical points are found in Appendix~\ref{app:BKT}.) The unconstrained model $H_K'$ would realize two decoupled Kitaev chains on even and odd sites. The constrained model $\mathcal P H_K' \mathcal P$, however, prefers to keep every other site (nearly) empty; the remaining sites are now unconstrained.

This is a general mechanism for realizing any unconstrained phase in the constrained Hilbert space: simply leave every other site empty. Then \emph{any} state on the remaining sites trivially satisfies the constraint.

Secondly, we introduce bond-alternation to $H_\textrm{hard}$ at $g=0$:
\begin{equation}
H_\delta = - \sum_i (1-(-1)^i\delta)\left( P_{i-1} c^\dagger_i c^{\vphantom \dagger}_{i+1}  P_{i+2} + \textrm{h.c.} \right). \label{eq:dimer}
\end{equation}
We know that the LL is stable for small enough $\delta$ (see Fig.~\ref{fig:PXYP}(b)), but Fig.~\ref{fig:phasediagramconstrained}(b) shows that it persists all the way to $|\delta| \approx 0.9$. At that point, there is a Pokrovsky-Talapov transition \cite{Pokrovsky80} into a commensurate phase with $1/4$ filling (see Appendix~\ref{app:PT} for details). For positive $\delta$, the two ground states are symmetry protected topological (SPT) phases with respect to the remaining spatial inversion symmetry, one of which is non-trivial with protected degeneracies in its entanglement spectrum. There are no physical edge modes since inversion is broken near the boundary.

Thus, in addition to verifying the stability of the Luttinger liquid, we found various symmetry-breaking and topological phases in the constrained Hilbert space.

\textbf{\emph{Conclusion.}}---In this work we have elucidated how soft and hard constraints in one-dimensional systems can give rise to emergent $U(1)$ symmetry and gaplessness without fine-tuning. Experimentally verifying and exploring these gapless phases---including the longer-range constraints---is within reach \cite{Keesling18}.

A natural question is whether the same principle might also help to stabilize, e.g., Fermi liquids in higher dimensions. Or staying in one spatial dimension, one could explore larger symmetry groups and, relatedly, central charges $c>1$.

An important open problem is the classification of phases in the constrained Hilbert space. Above, we have discussed how any unconstrained phase can be trivially realized in the constrained Hilbert space by breaking translation symmetry. It is natural to explore realizations that preserve translation symmetry. More exotically, might there be phases that can only be realized and stabilized in constrained Hilbert spaces---much like how enforcing symmetry vastly expanded the zoo of one-dimensional phases of matter \cite{Schnyder08,Kitaev09,Ryu10,Fidkowski11class,Turner11class,Chen10,Schuch11}? A recent work answers a resounding and exciting \emph{yes} to the latter question for the case of gapless phases \cite{Jones19}.

\textbf{\emph{Acknowledgements.}}---The authors would like to thank Dave Aasen, Henrik Dreyer, Nicola Pancotti, Maksym Serbyn and Ryan Thorngren for stimulating conversations. RV is indebted to Nick G. Jones for his insights into Luttinger liquids and for a careful reading of the manuscript. RV was supported by the German Research Foundation (DFG) through the Collaborative Research Center SFB 1143, AV is supported by a Simons Investigator Award and by the AFOSR MURI grant FA9550- 14-1-0035, and FP acknowledges the support of the DFG Research Unit FOR 1807 through grants no. PO 1370/2-1, TRR80, the European Research Council (ERC) under the European Union's Horizon 2020 research and innovation program (grant agreement no. 771537), and the Deutsche Forschungsgemeinschaft (DFG, German Research Foundation) under Germany's Excellence Strategy - EXC-2111 - 390814868.

\bibliography{arxiv.bbl}

\pagebreak

\widetext
\ifx\targetformat\undefined
\begin{center}
	\textbf{\large Supplemental Materials}
\end{center}

\setcounter{equation}{0}
\setcounter{figure}{0}
\setcounter{table}{0}
\makeatletter
\renewcommand{\theequation}{S\arabic{equation}}
\renewcommand{\thefigure}{S\arabic{figure}}
\renewcommand{\bibnumfmt}[1]{[S#1]}
\renewcommand{\citenumfont}[1]{S#1}

\else
\appendix
\setcounter{equation}{0}
\setcounter{figure}{0}
\setcounter{table}{0}
\renewcommand{\theequation}{A\arabic{equation}}
\renewcommand{\thefigure}{A\arabic{figure}}
\renewcommand{\bibnumfmt}[1]{[A#1]}
\fi

\section{Fermionic and spin models \label{app:spin}}

For clarity and completeness, here we include all the hardcore bosonic/spin models that are related to the fermionic models of the main text via the Jordan-Wigner transformation:
\begin{align}
\sigma^+_i &\equiv \exp\bigg(i \pi \sum_{j<i} n_j \bigg) c^\dagger_i, \label{eq:JW1} \\
\sigma^-_i &\equiv \exp\bigg(i \pi \sum_{j<i} n_j \bigg) c_i,  \label{eq:JW2}  \\
Z_i &\equiv 2n_i-1.  \label{eq:JW3} 
\end{align}
(We will use $X,Y,Z$ to represent the spin-$1/2$ Pauli matrices, i.e. $\sigma^\pm_i =\frac{1}{2} \left( X_i \pm i Y_i \right)$.) In particular, the Kitaev chain, $H_K = - \sum_i \left( c_i^\dagger c^{\vphantom \dagger}_{i+1} + c_i^\dagger c_{i+1}^\dagger + \textrm{h.c.} \right)$, takes the \emph{same} form in terms of the hardcore bosons defined by the above Jordan-Wigner transformation: $H_K =  -\sum_i \left( \sigma^+_i\sigma^-_{i+1} + \sigma^+_i \sigma^+_{i+1} + \textrm{h.c.} \right) =  - \sum_i X_i X_{i+1}$, known as the Ising chain. 

\subsection{The Kitaev/Ising chain with a soft constraint}

In the main text we considered the following fermionic model:
\begin{align}
H_\textrm{soft} &=H_K+ 2g \sum_i n_i + U \sum_i n_i n_{i+1} \notag \\
&=  - \sum_i \left( c_i^\dagger c^{\vphantom \dagger}_{i+1} + c_i^\dagger c_{i+1}^\dagger + \textrm{h.c.} \right) + 2g \sum_i n_i + U \sum_i n_i n_{i+1}. \label{eq:Ham1f}
\end{align}

The corresponding spin-$1/2$ model (ignoring overall additive constants) is:
\begin{equation}
H_\textrm{soft} = \sum_i \left(-  X_i X_{i+1} +  g \; Z_i  + U \; n_i n_{i+1} \right). \label{eq:Ham1s}
\end{equation}

\subsection{The Kitaev/Ising chain with a hard constraint}

Taking the $U\to \infty$ limit of the above model, we obtained the following effective model in the constrained Hilbert space:
\begin{equation}
H_\textrm{hard} = - \sum_i P_{i-1} \left( c_i^\dagger c^{\vphantom \dagger}_{i+1} + \textrm{h.c.} \right) P_{i+2} + 2 g \sum_i n_i . \label{eq:Hamconstrainedf}
\end{equation}

The corresponding spin-$1/2$ chain is:
\begin{equation}
H_\textrm{hard} =  \sum_i \left( - P_{i-1} \frac{ X_i X_{i+1} + Y_i Y_{i+1} }{2}  P_{i+2} +g Z_i \right) . \label{eq:Hamconstraineds}
\end{equation}
Here $P_i = (1-Z_i)/2$ projects onto a down spin.

At next order in $1/U$, the fermionic model has its $U(1)$ symmetry explicitly broken by $- \frac{2}{U} \sum_i \left( P_{i-2} c^\dagger_{i-1}  P_i  c^\dagger_{i+1} P_{i+2} + \textrm{h.c.}\right)$. In the spin model, this is $\frac{1}{U} \sum_i  P_{i-2}  P_i P_{i+2} \left( Y_{i-1} Y_{i+1}  -X_{i-1} X_{i+1} \right)$.

\subsection{The free fermion/XY chain with a soft constraint}

The aforementioned effective model had an explicit $U(1)$ symmetry. In fact, it also arises as the hard-constraint limit of the following $U(1)$ symmetric fermionic model, as considered in the main text:
\begin{equation}
H_\textrm{soft}^{U(1)} = - \sum_i \left( c^\dagger_i c^{\vphantom \dagger}_{i+1} + \textrm{h.c.} \right) + 2g \sum_i n_i + U \sum_i n_i n_{i+1}. \label{eq:U1softf}
\end{equation}

The corresponding spin-$1/2$ chain is:
\begin{equation}
H_\textrm{soft}^{U(1)} =  \sum_i \left(-  \frac{X_i X_{i+1} + Y_i Y_{i+1}}{2} +  g \; Z_i  + U \; n_i n_{i+1} \right). \label{eq:U1softs}
\end{equation}
Writing $n_i = (1+Z_i)/2$, one can see that the latter is equivalent to the XXZ chain with $\Delta = U/2$ and $h = g + U/2$.

\subsection{$U(1)$-breaking and dimerization in the constrained limit}

\subsubsection{$U(1)$-breaking}

\begin{align}
H_\lambda &=  (1-\lambda) \; H_\textrm{hard}(g) - \lambda \sum_i P_{i-2} P_i P_{i+2} \left( c^\dagger_{i-1} c^{\vphantom \dagger}_{i+1} + c^\dagger_{i-1} c^\dagger_{i+1}  + \textrm{h.c.} \right) \label{eq:U1breakf} \\
&= (1-\lambda) \; H_\textrm{hard}(g) - \lambda \sum_i P_{i-2} X_{i-1} P_i X_{i+1} P_{i+2}. \label{eq:U1breaks}
\end{align}

\subsubsection{dimerization}

\begin{align}
H_\delta &= - \sum_i (1-(-1)^i\delta)\left( P_{i-1} c^\dagger_i c^{\vphantom \dagger}_{i+1}  P_{i+2} + \textrm{h.c.} \right) \label{eq:dimerf} \\
&= - \sum_i (1-(-1)^i\delta) P_{i-1} \frac{X_i X_{i+1} + Y_i Y_{i+1}}{2}P_{i+2}. \label{eq:dimers}
\end{align}

\section{Relation to the axial/anisotropic next-nearest-neighbor Ising (ANNNI) model \label{app:ANNNI}}

\begin{figure}[h]
\includegraphics{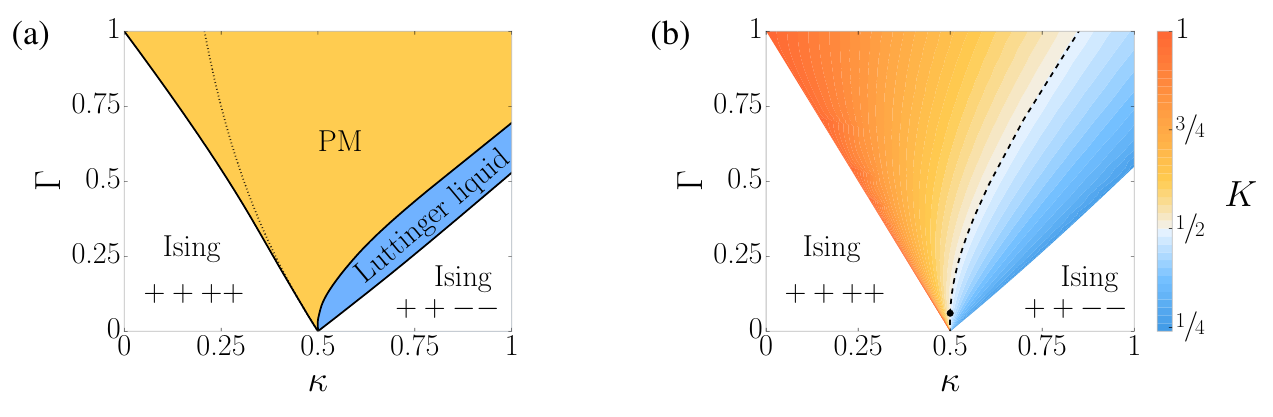}
\caption{(a) Phase diagram of the ANNNI model (obtained by mapping from Fig.~1 of the main text). The dotted line is the exactly-solvable Peschel-Emery line $\Gamma = \frac{1}{4\kappa} - \kappa$ along which the ground state is a matrix product state with bond dimension $\chi=2$ and $\xi = - \frac{1}{\ln(2 \kappa)} $ \cite{Peschel81}. (b) Phase diagram of the $U(1)$-enhanced ANNNI model, dual to Fig.~3 of the main text. Note that the values of the Luttinger liquid parameter are simply those of Fig.~3 of the main text, and depending one's convention one would give them a different value after the Kramers-Wannier transformation, especially considering that the $U(1)$ symmetry is no longer on-site. The dashed line denotes $K=1/2$ and the black dot signals where this line intersects $\kappa = 1/2$. \label{fig:ANNNI}}
\end{figure}

We can rewrite the model $H_\textrm{soft}$ appearing in the main text as follows (also see Eq.~\eqref{eq:Ham1s}):
\begin{align}
H &= \sum_i \left(-  X_i X_{i+1} +  g \; Z_i  + U \; n_i n_{i+1} \right) = \sum_i \left( -X_i X_{i+1} + g Z_i + \frac{U}{4} (Z_i+1)(Z_{i+1} + 1) \right) \\
&= \sum_i \left( - X_i X_{i+1} + \left(g + \frac{U}{2} \right) Z_i + \frac{U}{4} Z_i Z_{i+1} \right) + \textrm{cst}
\end{align}
If we drop the constant and perform a Kramers-Wannier transformation,
\begin{equation}
\begin{array}{ccl}
\tilde X_i &\equiv & (-1)^i \cdots Z_{i-2} Z_{i-1} Z_i \\
\tilde Z_i &\equiv & X_i X_{i+1},
\end{array}
\end{equation}
then we obtain
\begin{align}
H &= \sum_i \left( - \tilde Z_i - \left( g + \frac{U}{2} \right) \tilde X_i \tilde X_{i+1} + \frac{U}{4} \tilde X_i \tilde X_{i+2}  \right) \\
&\propto \sum_i \left(- \tilde  X_i \tilde X_{i+1} - \Gamma \tilde Z_i + \kappa \tilde X_i \tilde X_{i+2}  \right)  \qquad \textrm{with} \quad
\left\{
\begin{array}{ccl}
\Gamma &= & \left( g + \sfrac{U}{2} \right)^{-1} \\
\kappa &= & \frac{1}{2} (1 + \sfrac{2g}{U})^{-1} = \frac{U}{4} \Gamma.
\end{array}
\right.
\end{align}
We thus see that the model is equivalent to the ANNNI model (where we have used its traditional parametrization). This mapping allows us to plot the phase diagram in Fig.~1 (of the main text) in terms of $\kappa$ and $\Gamma$. This is shown in Fig.~\ref{fig:ANNNI}(a). Moreover, we show the $U(1)$-enhanced model in Fig.~\ref{fig:ANNNI}(b), which is the analogue of Fig.~3 of the main text. (Explicitly, the integrable $U(1)$-enhanced model is $H = \sum_i \left( -\hat X_i \hat X_{i+1} + \frac{\Gamma}{2}(\tilde X_{i-1} \tilde Z_i \tilde X_{i+1} - \tilde Z_i ) + \kappa \tilde X_i \tilde X_{i+2} \right)$, where the generator of the $U(1)$ symmetry is $\propto \sum_i \tilde X_i \tilde X_{i+1}$.)

Despite the (soft-)constrained model in the main text mapping to the ANNNI model, there are two important differences in the representation of these models, with distinct physical consequences.

Firstly, they are related by a non-local (Kramers-Wannier) transformation. Hence, whereas the model in the main text has emergent $U(1)$ symmetries whose charges are \emph{local}, the gapless phase in the ANNNI model has a non-local $U(1)$ symmetry. (More precisely, its generator is an matrix product operator, and operators with the smallest possible charge are non-local string operators.) Conceptually, this makes the model in the main text (and its phenomenology, such a stable gapless phases) arguably easier to understand. Quantitatively, it for example affects the oscillations of the correlation functions: in the model in the main text, the $\langle X_i X_j \rangle$ correlation functions do not have (sign-changing) oscillations. For the corresponding ANNNI model, however, the $\langle \tilde X_i \tilde X_j \rangle$ correlations (in addition to having a different scaling dimension), pick up large oscillations determined by the $k_F$ wavevector. (Relatedly, the values of $K$ shown in Fig.~\ref{fig:ANNNI}(b) need to be interpreted with care: strictly speaking, they refer to the values of the Luttinger liquid parameter \emph{prior} to the Kramers-Wannnier transformation; see Fig.~3 of the main text.)

Secondly, the parameters in the Hamiltonian are also non-locally related. In particular, the interesting and integrable limit $U \to \infty$ with $g$ finite, is mapped to a \emph{single} point of the ANNNI phase diagram: $\Gamma=0$ and $\kappa = \sfrac{1}{2}$. This is yet another way in which the model of the main text is easier to analyze than the ANNNI model. In particular, there are two hotly-debated questions about the ANNNI model which are straightforwardly addressed in terms of the model of the main text:
\begin{enumerate}
	\item Does the gapless phase of the ANNNI model have a finite extent at $\kappa=\sfrac{1}{2}$? This is equivalent to asking whether the model of the main text is gapless for $g= 0$ and $U$ large. We answered this in the positive by first using Bethe ansatz to show that the $U\to \infty$ limit is gapless, and subsequently arguing that this is stable for finite (but large) $U$ due to $K<\sfrac{1}{2}$. Moreover, we numerically found that for $g=0$, the model was gapless for $U \gtrapprox 50$; this means that the ANNNI model is gapless for $\kappa = \sfrac{1}{2}$ and $\Gamma \lessapprox 0.04$.
	\item Does the PM phase reach $\Gamma \to 0$? A perturbative calculation around the $\Gamma = 0$ and $\kappa = 1/2$ limit seemingly proved that the PM phase terminates at a finite value of $\Gamma$ \cite{Chandra07b}. However, subsequent DMRG\cite{Beccaria07} and iDMRG\cite{Nagy11} works could not corroborate this. Indeed, an earlier analytical work by Peschel and Emery already showed that there is an exactly solvable line in the PM phase that terminates at the $\Gamma = 0$ and $\kappa = 1/2$ point \cite{Peschel81} (dotted line in Fig.~\ref{fig:ANNNI}(a)).
	
	We now resolve this paradox. The claim in Ref.~\cite{Chandra07b} was based on finding that there are only two critical emanating from the $\Gamma = 0$ and $\kappa = 1/2$ point at leading-order in $\Gamma$. But at this order, the system develops an explicit $U(1)$ symmetry; at this level Fig.~\ref{fig:ANNNI}(a) and Fig.~\ref{fig:ANNNI}(b) are indistinguishable, and the latter indeed only has two critical lines emerging from this point. The catch is that part of the gapless phase opens up a gap at the next-to-leading order in $\Gamma$, becoming the PM phase. (We have discussed this in detail in the main text, relating it to the value of the Luttinger liquid parameter.) In conclusion, there are three critical lines (as shown in Fig.~\ref{fig:ANNNI}(a)), with the PM phase reaching the multi-phase point.
\end{enumerate}

Lastly, we would be remiss not touching upon the early seminal work on the two-dimensional classical ANNNI model. In particular, in 1981, Villain and Bak realized that near $\kappa \to 1/2$ and at low temperature (related to small $\Gamma$), the systems develops an effective $U(1)$ symmetry \cite{Villain81}. Moreover, in that same work, they used a free-fermion approach where they derived formulas for exponents in the algebraic phase which are in close agreement with those quoted in our main text. We can now interpret this as Villain and Bak having realized the solvability of the (hard-)constrained model, apparently almost two decades before it was explicitly discovered \cite{Alcaraz99}. Nevertheless, it is not entirely clear to what extent these approaches coincide, especially since the (hard-)constrained model is \emph{not} a free-fermion model (albeit a projection of one). It could be interesting to explore these relationships further. Relatedly, note that the two-dimensional classical and one-dimensional quantum model are not trivially related. Indeed, whereas the gapless phase of the 1D quantum ANNNI model was established early on, it had been claimed that in the 2D classical model this phase might be of vanishing width \cite{Shirahata01,Derian06,Chandra07}. This issue was only settled in recents years, concluding that there is an extended phase \cite{Shirakura14,Matsubara17}.

\section{Perturbation theory \label{app:pert}}

Here we calculate the effective Hamiltonian for the Kitaev/Ising chain with a soft constraint $\sim U$ in the limit where $U \to \infty$.

We set up the perturbation theory (in the bosonic language) as follows: $H = H_0 + V$ with $H_0 = U \sum_i n_i n_{i+1}$ and $V = \sum_i \left( - X_i X_{i+1} + g Z_i \right)$. Let $\mathcal P$ be the projector onto the constrained Hilbert space.

For large $U$, we can write down an effective Hamiltonian: $H_\textrm{eff} = \sum_{k=0}^\infty \frac{1}{U^k} H_\textrm{eff}^{(k)} $. For $U \to \infty$, we have $H_\textrm{eff} = H_\textrm{eff}^{(0)}$. We know $H_\textrm{eff}^{(0)} = \mathcal P V \mathcal P$, which gives us
\begin{equation}
H_\textrm{eff}^{(0)}= \sum_i \left( - P_{i-1} \sigma^+_i \sigma^-_{i+1}P_{i+2} + h.c. + g Z_i \right).
\end{equation}

We also know that $H_\textrm{eff}^{(1)} = \mathcal P V G V \mathcal P$, where $G = \mathcal Q \frac{U}{E_0 - H_0} \mathcal Q$, with $\mathcal Q = 1- \mathcal P$ and $E_0 = \mathcal P H_0 \mathcal P = 0$ \cite{Kato49,Takahashi77}. After some algebra, one obtains
\begin{align}
H_\textrm{eff}^{(1)}= \sum_i \bigg( &- \frac{1}{2} P_{i-2} \sigma^+_{i-1} P_i \sigma^-_{i+1} P_{i+2} - P_{i-2} \sigma^+_{i-1} \sigma^-_{i} \sigma^+_{i+1} \sigma^-_{i+2} P_{i+3} - 2 P_{i-2} \sigma^+_{i-1} P_i \sigma^+_{i+1} P_{i+2} + h.c. \\
& + 3 n_i - 3n_i n_{i+2} - \frac{1}{3} n_i n_{i+3} - L \bigg).
\end{align}
Note that the first two terms are $U(1)$-preserving: the first is longer-range hopping, the second is pair hopping. The last term on the first line is the $U(1)$-breaking term.

\section{Luttinger liquid parameter in the XXZ chain \label{app:XXZ}}

We start from the model $H^{U(1)}_\textrm{soft}$ of the main text, parameterizing it in terms of $\Delta = \sfrac{U}{2}$ and $h = g + \sfrac{U}{2}$. Using the Jordan-Wigner transformation (see Eqs.~\eqref{eq:JW1}--\eqref{eq:JW3}), we map it to the XXZ chain:
\begin{align}
H &= - \sum_i \left( c^\dagger_i c^{\vphantom \dagger}_{i+1} + \textrm{h.c.} \right) + 2 \Delta \sum_i n_i n_{i+1} + 2(h-\Delta) \sum_i n_i \\
&= \frac{1}{2} \sum_i \left( X_i X_{i+1} + Y_i Y_{i+1} + \Delta Z_i Z_{i+1} + 2h Z_i \right) + \textrm{cst}.
\end{align}

\begin{figure}[h]
\includegraphics[scale=0.25]{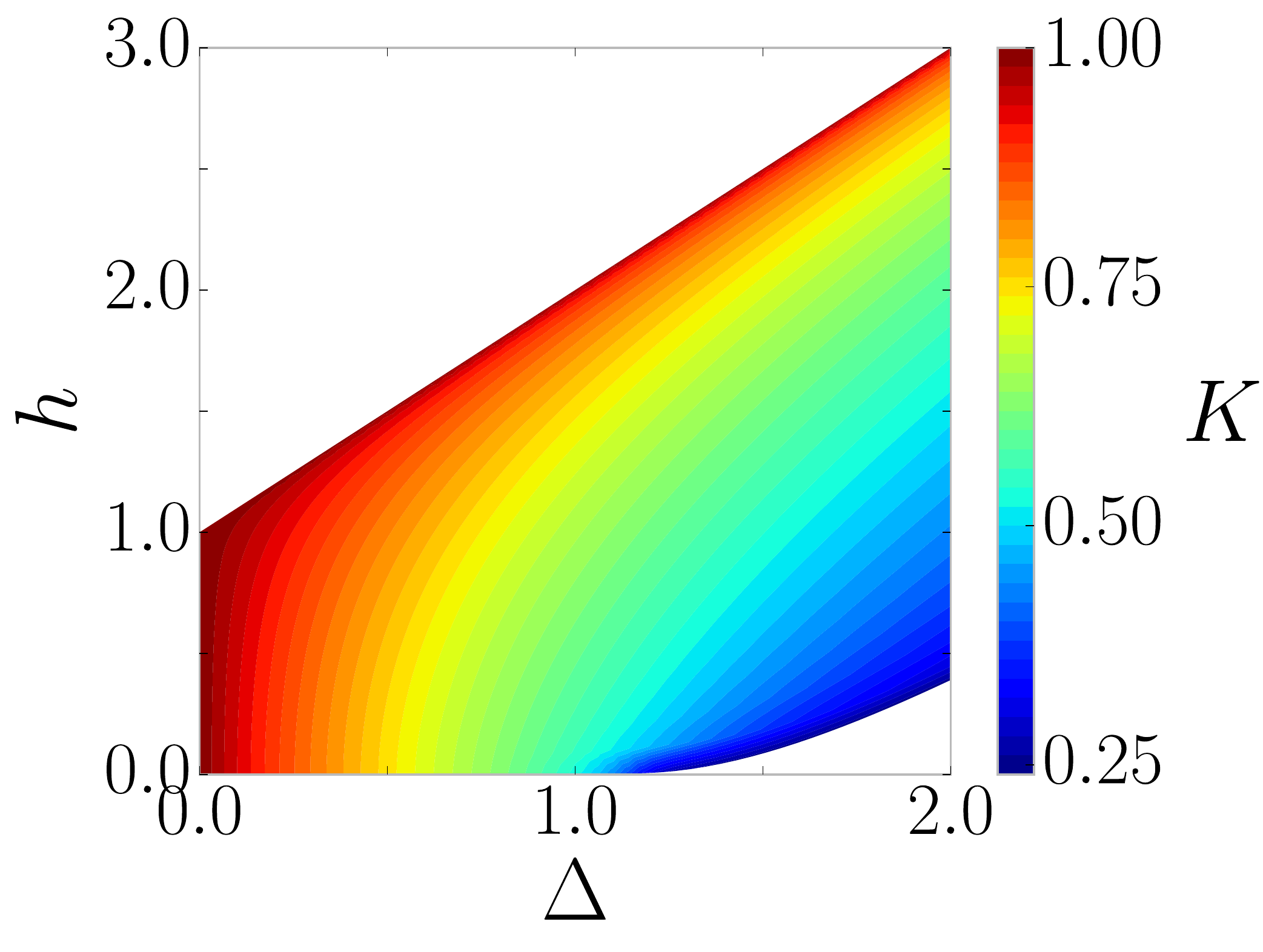}
\caption{Luttinger liquid parameter for the XXZ chain. \label{fig:XXZ}}
\end{figure}

The integral equations which specify its Luttinger liquid parameter were derived in Ref.~\cite{Bogoliubov86}. By numerically solving these, we obtain the data in Fig.~\ref{fig:XXZ}. For completeness, we reproduce the integral equations here, following the notation of Ref.~\cite{Cabra98}. This will also allow us to derive simple analytic relationships in the constrained limit in the next section. In particular, we present the formulas for $\Delta > 1$. (For $\Delta < 1$, the same formulas apply with all trigonometric functions replaced by their hyperbolic counterparts and vice versa.)

We consider $\Delta = \cosh \gamma$ and the filling $\rho_0$ to be fixed. (We will determine the corresponding field $h$.) We then define the following two kernels:
\begin{equation}
\mathcal K(\eta) \equiv \frac{\tanh \gamma }{\tanh^2 \gamma \cos^2(\eta/2) + \sin^2(\eta/2)} \qquad \textrm{and} \qquad \mathcal G(\eta) \equiv \frac{\coth \gamma }{\coth^2 \gamma \sin^2(\eta/2) + \cos^2(\eta/2)}. \label{eq:kernels}
\end{equation}
In terms of these, we define $\sigma(\eta)$ and $\Lambda$ by the following two integral equations:
\begin{equation}
2\pi \sigma(\eta) \equiv \mathcal G(\eta) - \int_{-\Lambda}^{\Lambda} \mathcal K(\eta - \eta') \sigma(\eta') \mathrm d \eta '
\qquad \textrm{and} \qquad
\int_{-\Lambda}^{\Lambda } \sigma(\eta) \mathrm d \eta \equiv \rho_0.
\end{equation}
The Luttinger liquid parameter $K$ is given by $K = \xi(\Lambda)^2$ where $\xi(\eta) \equiv 1 - \frac{1}{2\pi} \int_{-\Lambda}^{\Lambda} \mathcal K(\eta - \eta') \xi(\eta') \mathrm d \eta'$. The field is determined by $h = \varepsilon(\Lambda)/\sqrt{K}$, where $\varepsilon(\eta) = \frac{\sinh^2 \gamma}{\cosh \gamma - \cos \eta } - \frac{1}{2\pi} \int_{-\Lambda}^{\Lambda} \mathcal K(\eta-\eta') \varepsilon(\eta')\mathrm d \eta'$.

\section{Analytic relationships in the \texorpdfstring{$U\to \infty$}{} limit \label{app:XXZ2}}

As $\Delta \to +\infty$, we see from Eq.~\eqref{eq:kernels} that $\lim_{\Delta \to + \infty} \mathcal K(\eta) = \lim_{\Delta \to + \infty} \mathcal G(\eta) = 1$. Hence, $2\pi\sigma(\eta) = 1 - \int_{-\Lambda}^{\Lambda} \sigma(\eta ') \mathrm d \eta '$, whose solution is $\sigma(\eta) = \frac{1}{2\pi + 2 \Lambda}$. In particular, $\sigma(\eta)$ is constant, such that the condition $\int_{-\Lambda}^{\Lambda } \sigma(\eta) \mathrm d \eta \equiv \rho_0$ implies that $\sigma(\eta) = \frac{\rho_0}{2\Lambda}$. Equating these two solutions gives us $\Lambda = \frac{\pi\rho_0}{1-\rho_0}$.

Similarly, $\xi(\eta)$ is also constant, with the solution from the integral equation being $\xi(\eta) = \frac{1}{1+\Lambda/\pi} = 1 - \rho_0$. We conclude that the Luttinger liquid parameter $K= (1-\rho_0)^2$, as claimed in the main text. (It will also be useful to note that $\Lambda = \frac{\pi}{\sqrt{K}} - \pi$. Note that for $K: 1/4 \to 1$, we have $\Lambda:\pi \to 0$.)

To determine the field $h$, we note that for large $\gamma$, we can approximate the integral equation for $\varepsilon(\eta)$ as $\varepsilon(\eta) = \Delta + \cos \eta - \frac{1}{2\pi} \int_{-\Lambda}^{\Lambda} \varepsilon(\eta') \mathrm d \eta'$ where we used that $\frac{\sinh^2 \gamma}{\cosh \gamma - \cos \eta } \approx \frac{\Delta^2}{\Delta - \cos \eta} = \Delta \left( 1-\frac{\cos \eta}{\Delta} \right)^{-1} \approx \Delta + \cos \eta$. Integrating both sides of the integral equation and defining $\mathcal I \equiv \int_{-\Lambda}^{\Lambda} \varepsilon(\eta) \mathrm d \eta$, we obtain $\mathcal I = 2 \Lambda \Delta + 2 \sin \Lambda - \frac{\Lambda}{\pi} \mathcal I$, i.e. $\mathcal I = \frac{2 \Lambda \Delta + 2 \sin \Lambda}{1 + \Lambda/\pi}$. Thus,
\begin{equation}
\varepsilon(\eta) = \Delta + \cos \eta - \frac{\mathcal I}{2\pi} = \Delta + \cos \eta - \frac{\Lambda \Delta + \sin \Lambda}{\pi + \Lambda} = (1- \rho_0) \Delta + \cos \eta - \frac{1-\rho_0}{\pi} \sin \Lambda.
\end{equation}
In conclusion, we derive that
\begin{equation}
h - \Delta = \frac{\varepsilon(\Lambda)}{\sqrt{K}} - \Delta = \frac{1}{\sqrt{K}}\cos \Lambda - \frac{1}{\pi} \sin \Lambda = \frac{1}{\pi} \sin \left( \frac{\pi}{\sqrt{K}} \right) - \frac{1}{\sqrt{K}} \cos \left( \frac{\pi}{\sqrt{K}} \right).
\end{equation}

Given the identification of the models in the main text, i.e. $U = 2 \Delta$ and $g = h- \Delta$, we conclude that in the $U \to +\infty$ limit, we have $\pi g = \big( \sin(x) - x \cos(x) \big) \big|_{x = \pi / \sqrt{K}}$, in agreement with what was claimed in the main text.

\section{Criticality in constrained model with \texorpdfstring{$U(1)$}{}-breaking terms \label{app:BKT}}

Here we analyze $H_\lambda$ of the main text, setting $g=-1$:
\begin{equation}
H_\lambda = - (1-\lambda)\left[ \sum_i P_{i-1} \left( c_i^\dagger c^{\vphantom \dagger}_{i+1} + \textrm{h.c.} \right) P_{i+2} + 2  n_i \right] - \lambda \sum_i P_{i-2} P_i P_{i+2} \left( c^\dagger_{i-1} c^{\vphantom \dagger}_{i+1} + c^\dagger_{i-1} c^\dagger_{i+1}  + \textrm{h.c.} \right).
\end{equation}

\subsection{Berezinskii-Kosterlitz-Thouless (BKT) universality as \texorpdfstring{$\lambda \to \lambda_\textrm{BKT}$}{}}

\begin{figure}[h]
\includegraphics[scale=0.25]{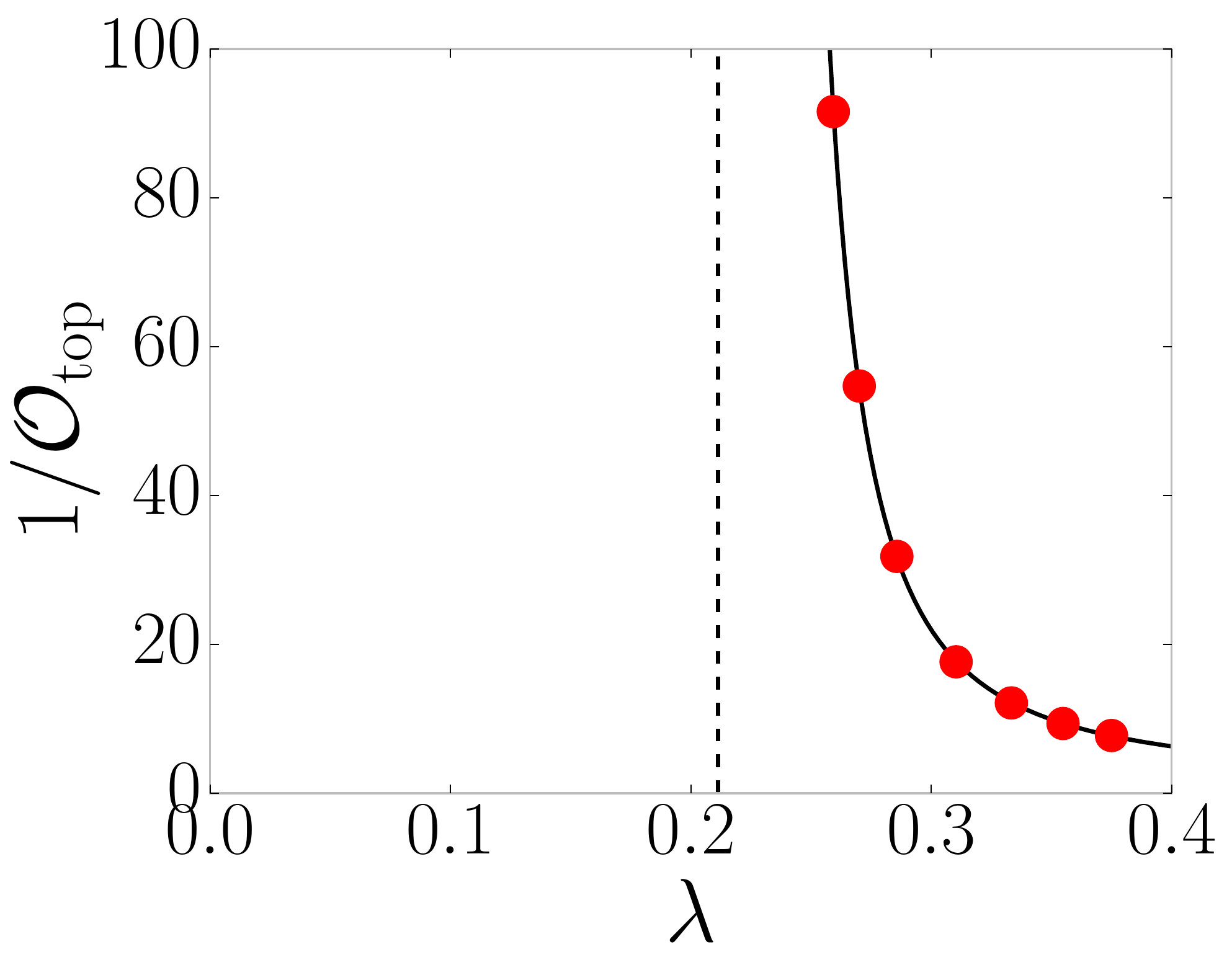}
\caption{BKT transition in $H_\lambda$ at $\lambda_\textrm{BKT} \approx 0.21$. \label{fig:BKTPXPXP}}
\end{figure}

From fitting how the order parameter goes to zero, using the BKT formula (see main text), we obtain $\lambda_\textrm{BKT} \approx 0.21$ (see Fig.~\ref{fig:BKTPXPXP}).

\subsection{Ising criticality as \texorpdfstring{$\lambda \to \lambda_\textrm{Ising}$}{}}

\begin{figure}[h]
\includegraphics{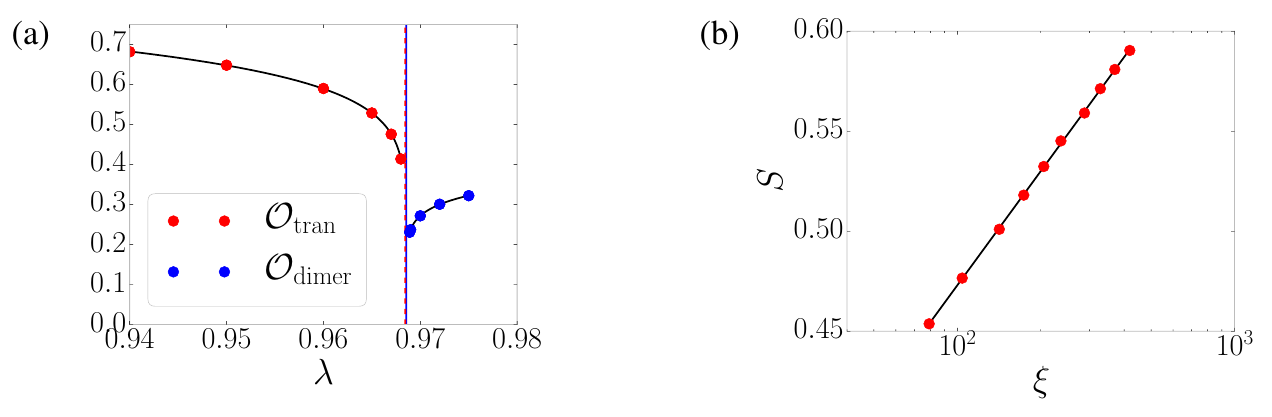}
\caption{An Ising transition in $H_\lambda$. (a) Locating the critical point at $\lambda_\textrm{Ising} \approx 0.9685$ and extracting scaling dimensions $\beta_\textrm{dimer} \approx 0.11$ and $\beta_\textrm{tran} \approx 0.12$ (b) Entanglement scaling $S(\chi) \sim \frac{c}{6} \ln \xi(\chi)$ at $\lambda = 0.96846$ to extract the central charge $c\approx 0.494$. \label{fig:PXPXP2}}
\end{figure}

We first locate the critical point by looking at order parameters. For large $\lambda$, the state spontaneously breaks translation symmetry, for which we can define the order parameter $\mathcal O_\textrm{dimer} = \langle n_1 - n_0 \rangle$, which vanishes at the critical point as shown in Fig.~\ref{fig:PXPXP2}(a). There are only very few points accessible (given the narrow window of the phase), from which we extract an approximate scaling dimension $\beta_\textrm{dimer} \approx 0.11$.

\begin{figure}[h]
\includegraphics[scale=1]{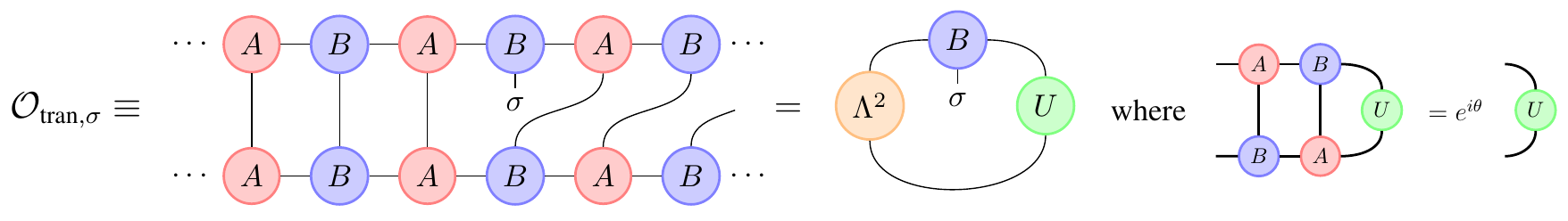}
\caption{We define the (dis)order parameter $\mathcal O_\textrm{tran} = \sqrt{ \sum_{\sigma = \uparrow,\downarrow} |\mathcal O_{\textrm{tran},\sigma}|^2 }$ for the disordered (topological) phase of $H_\lambda$ and we calculate an efficient expression in terms of the tensors defining the Matrix Product State (MPS). The MPS is in right-canonical form, i.e. the transfer matrix has the identity as its dominant right eigenvector, and $\Lambda^2$ as its dominant left eigenvector. \label{fig:Otran}}
\end{figure}

Considering that the phase is much more stable when coming from the other side, we would like to look at the expectation value of the corresponding disorder operator (i.e. the operator which can be regarded as being condensed in the symmetry-preserving phase). For this we choose the operator which implements a shift on half the system, as shown in Fig.~\ref{fig:Otran}. This vanishes at the same point as shown in Fig.~\ref{fig:PXPXP2}(a), which gives the reliable value $\lambda_\textrm{Ising} \approx 0.9685$. The extracted critical exponent is $\beta_\textrm{tran} \approx  0.12$.

The values of the critical exponents, $\beta_\textrm{dimer} \approx 0.11$ and $\beta_\textrm{tran} \approx 0.12$ (the latter being more reliable than the former), are approximately consistent with the CFT prediction for the Ising universality class, where $\beta = 1/8 = 0.125$ for both the order and disorder parameters. In addition, we extract the central charge from entanglement scaling, shown in Fig.~\ref{fig:PXPXP2}(b). This gives a fit $c \approx 0.494$, in close agreement with the predicted value $c=1/2$.

\section{Criticality in constrained model with bond-alternation \label{app:PT}}

Here we analyze $H_\delta$ of the main text.

\subsection{Pokrovsky-Talapov (PT) criticality as \texorpdfstring{$\delta \to \delta_\textrm{PT}$}{}}

\begin{figure}[h]
\includegraphics[scale=1]{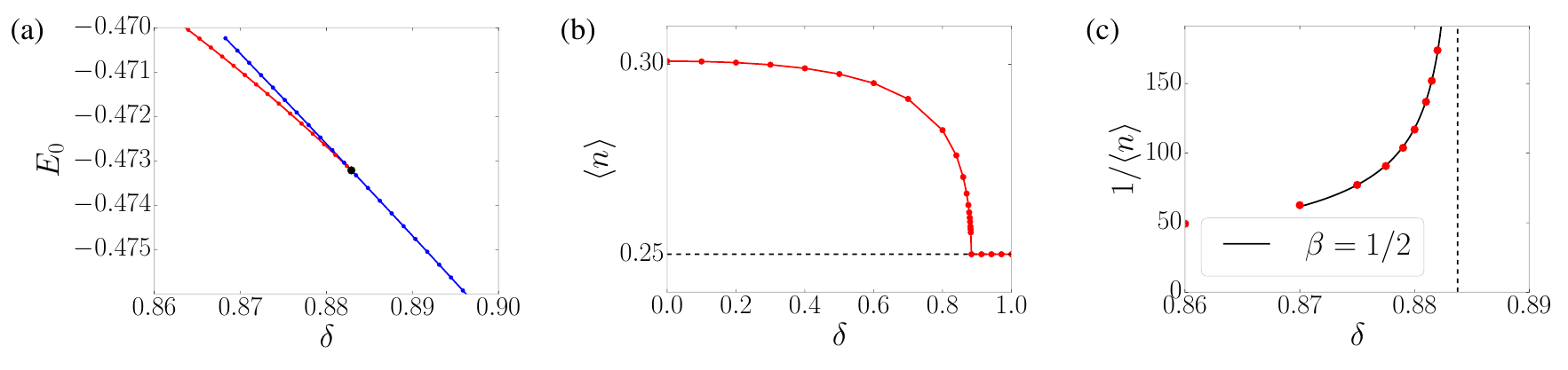}
\caption{Pokrovsky-Talapov criticality in $H_\delta$. (a) Ground state energy. There is a metastable state when coming from the right, which might lead one to misinterpret the transition as being first order. However, there is no kink. (b) Whilst the order parameter for $\delta> \delta_\textrm{PT}$ is discontinuous (not shown), observables evolve continuously when coming from the left, such as, e.g., the ground state filling. (c) The ground state filling approaches $1/4$ with critical exponent $\beta_\textrm{PT} = 1/2$. \label{fig:PT}}
\end{figure}

There is a PT transition at $\delta_\textrm{PT} \approx 0.883$. This is analzed in Fig.~\ref{fig:PT}. Note that this transition looks first order when coming from the right; this is also evidenced in Fig.~\ref{fig:PT}(a) where adiabatically following the ground state for $\delta > \delta_\textrm{PT}$ reaches a metastable state for $\delta < \delta_\textrm{PT}$. Nevertheless, the transition is continuous, and indeed the ground state energy from the left is tangent to the blue line in Fig.~\ref{fig:PT}(a).

Similarly, on the one hand, coming from $\delta > \delta_\textrm{PT}$, the order parameter has a discontinuity at the transition (not shown), whilst on the other hand, approaching from the left, physical observables will be continuous. This is shown in Fig.~\ref{fig:PT}(b), where we track the (incommensurate) ground state filling. In Fig.~\ref{fig:PT}(c) we confirm that this behavior is consistent with the critical exponent $\beta = 1/2$ predicted for the PT universality class.

\end{document}